\documentclass[a4paper,12pt,english]{article}

\usepackage{amsfonts,bm,amssymb,euscript,array,babel,cite}
\usepackage{amsmath,amsthm}
\usepackage[dvips]{epsfig}
\usepackage{slashed}



\newcommand{\be}{\begin{equation}} \newcommand{\ee}{\end{equation}}
\newcommand{\beq}{\begin{equation}} \newcommand{\eeq}{\end{equation}}
\newcommand{\beqa}{\begin{eqnarray}}
\newcommand{\eeqa}{\end{eqnarray}} 
\def\nn{\nonumber} \def\bea{\begin{eqnarray}} \def\eea{\end{eqnarray}}

\newcommand{\barr}{\begin{array}}
\newcommand{\earr}{\end{array}}

\def\a{\alpha}  
  \def\G{\Gamma}
 \def\d{\delta} 
 \def\e{\epsilon} 
    
 \def\L{\Lambda}  \def\m{\mu}
\def\n{\nu}

\def\R{{\mathbb R}}  
\def\Z{{\mathbb Z}} \def\one{\mbox{1 \kern-.59em {\rm l}}}

\def\bit{\begin{itemize}} \def\eit{\end{itemize}} 

\def\({\left(} \def\){\right)}

 \allowdisplaybreaks[3]

\begin{document}

\renewcommand{\title}[1]{\vspace{10mm}\noindent{\Large{\bf
#1}}\vspace{8mm}} \newcommand{\authors}[1]{\noindent{\large
#1}\vspace{5mm}} \newcommand{\address}[1]{{\itshape #1\vspace{2mm}}}

\begin{titlepage}

\begin{center}

\title{ \Large On Lie-algebraic solutions of the type IIB matrix model}

\vskip 3mm

\authors{Athanasios {Chatzistavrakidis
%${}^{1}$
}}

\vskip 3mm

\address{
%${}^1$ 
{\it Bethe Center for Theoretical Physics and Physikalisches Institut der Universit\"{a}t Bonn
\\
Nussallee 12, D-53115 Bonn, Germany} \\ \bigskip E-mail:
{than@th.physik.uni-bonn.de}}

\vskip 1.4cm

\textbf{Abstract}

\vskip 3mm

\begin{minipage}{14cm}%

\end{minipage}

\end{center}

A systematic search for Lie algebra solutions of the type IIB matrix model is performed. 
Our survey is based on the classification of all Lie algebras for dimensions up to five and of all 
nilpotent Lie algebras of dimension six. It is shown that Lie-type solutions of the equations of motion of the type IIB matrix model 
exist and they correspond to certain nilpotent and solvable Lie algebras. Their representation in terms of Hermitian matrices is discussed in detail. These algebras give rise to certain non-commutative spaces for which the corresponding star-products are provided.  Finally the issue of constructing quantized compact nilmanifolds and solvmanifolds 
based on the above algebras is addressed. 
%Furthermore we comment on the relation between these solutions and twisted tori backgrounds of the type IIB 
%superstring theory. 
%The correspondence between flux compactifications of type IIB superstring theory on 
%twisted tori and quantized nilmanifold solutions of the IKKT model supports the conjecture that the latter 
%provides a non-perturbative definition of the former. 

\end{titlepage}

\tableofcontents

%\newpage

\section{Introduction}

String-inspired Matrix Models (MM) were proposed as non-perturbative 
definitions of M theory \cite{Banks:1996vh} and type IIB superstring theory \cite{Ishibashi:1996xs}.
Such MM provide an interesting and simple framework to study the dynamics of branes, both 
analytically and numerically. 

Several solutions of the above MM have been identified. 
In particular, as far as the model of Ishibashi-Kawai-Kitazawa-Tsuchiya (IKKT) \cite{Ishibashi:1996xs}
%Banks-Fischler-Shenker-Susskind \cite{Banks:1996vh} 
is concerned,
%Several solutions of the model of Ishibashi-Kawai-Kitazawa-Tsuchiya (IKKT) \cite{Ishibashi:1996xs} are known as well. 
the first solutions appear in the original publication and they correspond to one or more 
%(parallel, 
%antiparallel)
 $D$-strings. Odd-dimensional $Dp$-brane solutions, in accord with type IIB superstring theory, were 
described and studied in \cite{Chepelev:1997ug,Fayyazuddin:1997zx,Fayyazuddin:1997yf,Aoki:1999vr,Aoki:1998vn}. On the 
other hand, compact non-commutative (NC) spaces, such as fuzzy tori, fuzzy spheres and other fuzzy homogeneous spaces, were 
shown to provide solutions upon adding extra terms (deformations) in the original action 
\cite{Myers:1999ps,Iso:2001mg,Kimura:2001uk,Kitazawa:2002xj,Azuma:2004zq,Steinacker:2003sd}. Compact 
solutions of the undeformed MM were described only recently \cite{Steinacker:2011wb}. The relation of the 
IKKT model to toroidal compactifications was already discussed from a different point of view in the 
pioneering paper \cite{Connes:1997cr}.

The fluctuations around solutions of the MM carry gauge degrees of freedom and provide a fruitful arena 
to study non-abelian gauge theories on NC space-time \cite{Aoki:1999vr}. Such backgrounds may therefore provide a natural set-up 
for model building, much like vacua of the type IIB string theory with $D$-branes. Such attempts were made in 
\cite{Aoki:2002jt,Aoki:2010gv,Grosse:2010zq}. More recently, configurations of intersecting NC branes in 
the IKKT model were studied and a realization of the gauge group and the chiral spectrum of the standard 
model was provided in \cite{Chatzistavrakidis:2011gs}. Similar considerations in a field-theoretical context 
were discussed in \cite{Aschieri:2006uw,Chatzistavrakidis:2009ix,Chatzistavrakidis:2010tq,Chatzistavrakidis:2010xi}.

In the present work we perform a survey on the possible Lie-type solutions to the (undeformed) IKKT MM. In other 
words, we examine which of the known and classified (semisimple, nilpotent and solvable) Lie algebras 
provide solutions to its equations of motion. Such an 
examination does not come without its restrictions. Indeed, the classification of Lie algebras beyond 
six dimensions becomes complicated. Therefore, our first restriction is to focus on all the Lie 
algebras of dimension up to five and all nilpotent Lie algebras of dimension six\footnote{The same 
classification was used in \cite{Kehagias:1994ys} in the construction of WZW models.}. For these low numbers of 
dimensions full classification tables exist \cite{mubarak1,morozov,Patera:1976ud} and therefore our task becomes tractable. 
Moreover, we shall be interested in the following only in non-abelian algebras. Abelian ones and algebraic sums of them are always solutions of the MM but they do not lead to interesting dynamics; therefore they will not be further discussed. 
%Finally, our survey focuses on solutions of the undeformed IKKT MM, i.e. without 
%the addition of any extra terms, and in particular on tree-level solutions. In fact, some algebras which are 
%not tree-level solutions might provide solutions at one loop. However, such cases will not be discussed in 
%the present work. 
Finally, let us mention that we work here with real Lie algebras; Lie algebras over 
different fields will not be considered.

Under the above requirements, scanning the classification tables of Lie algebras in section 3, we come up with the 
following result. There are only nine Lie-algebraic solutions to the IKKT MM, one of which is 3-dimensional, 
one 4-dimensional, three 5-dimensional and four 6-dimensional. Out of the above nine cases, seven correspond 
to nilpotent Lie algebras and two of them correspond to solvable ones. There are no semisimple Lie algebras 
providing tree-level solutions to the undeformed IKKT MM. 
%As a consequence, all the Lie algebras we find are
%non-compact. 
%This is in accord with previous results, where 
%it was shown that compact solutions based on semisimple Lie algebras may exist only upon addition of appropriate terms (deformations) in the model
% \cite{Iso:2001mg,Kimura:2001uk,Kitazawa:2002xj,Azuma:2004zq,Steinacker:2003sd} (see 
%however \cite{Steinacker:2011wb}).  

Having identified the Lie algebras which constitute solutions of the equations of motion of the model, the 
next step is to study 
whether they can be represented by Hermitian matrices. Evidently this is a necessary requirement 
in order for these Lie algebras to correspond to NC spaces which are indeed solutions of the MM. 
Utilizing the powerful results of Kirillov on the unitary representations of nilpotent Lie groups \cite{kirillov}, such representations are indeed determined for most of the relevant Lie algebras.

Finally, the issue of constructing compact non-commutative spaces based on these algebras is 
addressed. 
%Our interest mainly lies in 
%compact spaces. 
The main possibility which arises in this context is to consider spaces obtained as 
the quotient of a nilpotent Lie group by a compact discrete subgroup of it. Such spaces are known as 
nilmanifolds and they can be compact even when the nilpotent group is not \cite{malcev}. Some of them are known 
to provide string flux vacua based on the ideas of the seminal paper \cite{Scherk:1979zr}, see e.g. \cite{LopesCardoso:2002hd,Dall'Agata:2005ff,Hull:2005hk,Grana:2006kf} and references therein. 
%In section 4 we discuss the quantization of nilmanifolds 
%in the context of Weyl quantization and we provide the corresponding star products. 
%Although we cannot argue 
%that these compact NC nilmanifolds provide solutions to the IKKT model without further requirements.
%In order to perform this task, 
%further properties of these algebras have to be considered. In particular, one needs the corresponding 
%quadratic Casimir operators and invariant metrics. Essentially we find that only one out of the nine solutions 
%possesses a non-degenerate invariant metric with non-trivial quadratic Casimir operator. This algebra is a 
%6-dimensional nilpotent one and we proceed to the construction of the corresponding non-commutative space 
%based on it. 
%A discussion on the results and their relation 
%to known string vacua appears in section 5. 
In an appendix we collect some useful definitions on Lie algebras.

\section{The type IIB matrix model}

\paragraph{Action and symmetries.}

Let us briefly describe the IKKT or IIB matrix model, which was originally proposed in \cite{Ishibashi:1996xs} as a non-perturbative definition of the 
type IIB superstring theory. It is a 0-dimensional reduced matrix model defined by the action
\be
S  =  -{\L^4\over g^2}Tr({1\over 4}[X_a,X_b][X^a,X^b]
+{1\over 2}\bar{\psi}\Gamma ^a[X_a,\psi ]),
\label{IKKT-action}
\ee
where $X_a, a=0,\dots,9$ are ten hermitian matrices, 
and $\psi$ are sixteen-component Majorana-Weyl spinors of $SO(9,1)$.
Indices are raised and lowered with the invariant tensor $\eta_{ab}$, or 
possibly $\delta_{ab}$ in the Euclidean version of the model where $SO(9,1)$ is replaced by $SO(10)$.
The $\Gamma^a$ are generators of the corresponding Clifford algebra. 
$\L$ is an energy scale, which we will set 
% which later will be related to the scale of non-commutativity but for now we shall set it 
equal to one, $\L=1$, and work with dimensionless quantities. Finally, $g$ is a parameter which can be related to the 
gauge coupling constant.

The symmetry group of the above model contains the $U(N)$ gauge group
(where the limit $N \to \infty$ is understood) as well as the  $SO(10)$ or $SO(9,1)$ global symmetry.
Moreover, the model is ${\cal N}=2$ supersymmetric, with supersymmetries realized by the following transformations, 
\bea \d_{\e}\psi&=&\frac i2[X_a,X_b]\G^{ab}\e, \qquad
		\d_{\e}X_a=i\bar\e\G_a\psi, \\
			\d_{\xi}\psi&=&\xi, \qquad\qquad\qquad\quad ~ 
				\d_{\xi}X_a=0, \eea 
where $\Gamma^{ab}$ denotes the antisymmetrized product of gamma matrices as usual.
Therefore, the amount of supersymmetry indeed matches that of the type IIB superstring. 
Let us also note that the homogeneous $\e$-supersymmetry is inherited by the maximal ${\cal N}=1$ supersymmetry of 
super-Yang-Mills theory in ten dimensions.

It is important to stress that due to its 0-dimensional nature, the IKKT model is not defined on any predetermined space-time background. 
Instead, space-time emerges as a particular solution of the model, as we discuss in the following. This picture provides a dynamical origin for geometry and space-time.

\paragraph{Equations of motion and basic solutions.}

Varying the action (\ref{IKKT-action}) with respect to the matrices $X_a$ and setting $\psi=0$, the following equations of motion are obtained,
\be\label{ikkteoms} [X_b,[X^a,X^b]]=0. \ee 
Simple as they may appear, these equations admit diverse interesting and non-trivial solutions. 
Clearly, the simplest solution is given by a set of commuting matrices, $[X^a,X^b]=0$. In that case, 
the matrices $X^a$ can be simultaneously diagonalized and therefore they may be expressed as 
\be X^a = \mbox{diag}(X^a_1,X^a_2,\dots,X^a_N). \ee 
However, such solutions are in a sense degenerate and do not lead to interesting dynamics. In the following, commutative solutions will not be considered any further.

For notational convenience let us now split the ten matrices $X^a$ in two sets; we shall use the following notation,
\be\label{split} X^a=  \begin{pmatrix}
	  X^\mu \\
           X^{i}
\end{pmatrix}, \ee 
where the $X^{\m},\m=0,\dots,3$ correspond to the first four $X^a$ matrices and the $X^{i}, i=1,\dots,6$ to the six rest of the 
$X^a$ matrices respectively\footnote{Let us stress that although a splitting of the type $10=4+6$ is 
considered here, 
this is not a priori favoured by the dynamics of the model. For studies related to the 4-dimensionality of space-time in the IKKT model see e.g. \cite{Aoki:1998vn,Nishimura:2001sx,Kawai:2002ub}.}. 
In this notation, another solution of the equations (\ref{ikkteoms}) is given 
by
\be
 X^a = \begin{pmatrix}
	 \bar X^\mu \\
          0
\end{pmatrix} \ee 
where $\bar X^{\m}$ are the generators of the Moyal-Weyl quantum plane $\R^4_{\theta}$, which satisfy the commutation relation
\be 
[\bar X^\mu,\bar X^\nu] = i \theta^{\mu\nu},
\ee
where $\theta^{\m\n}$ is a constant antisymmetric tensor. This solution corresponds to a single non-commutative (NC) flat 3-brane, 
which corresponds to space-time emerging as a solution of the matrix model. 
Being a single brane, this solution is associated to an abelian gauge theory. An obvious generalization of the above solution is given by 
\be
 X^a = \begin{pmatrix}
	 \bar X^\mu \\
          0
\end{pmatrix}\otimes \one_{n},
\label{coinciding-branes}
\ee 
which is interpreted as $n$ coincident branes carrying a non-abelian $U(n)$ gauge theory.

\section{Lie-algebraic solutions}

In the present section we search for solutions of the IKKT model which have the structure of a Lie algebra. 
This task may be split in two steps. First the Lie algebra should solve 
%As it was mentioned in the introduction, we are looking for solutions of the undeformed model and at tree-level, 
%i.e. solutions of 
the equations of motion (\ref{ikkteoms}). Secondly, the algebras which pass the first test should possess 
representations in terms of Hermitian matrices, since only then they may be considered solutions of the 
IKKT model. We shall address these two issues separately below.

 The Lie algebras we study here are 
algebras over the field of real numbers. The classification we follow appears in the tables of \cite{Patera:1976ud}, which is a complete classification of real Lie algebras of dimension up to five and 
real nilpotent Lie algebras\footnote{For the definition of nilpotent and solvable Lie algebras the reader may 
consult the appendix.} of dimension six\footnote{Nilpotent algebras of dimension seven are also classified 
but they are not finitely many.}. Let us mention that a certain classification of solvable Lie algebras 
of dimension six was partially given in \cite{mubarak2} and later completed in \cite{turkowski} but here 
we shall restrict only on the cases mentioned above, appearing in \cite{Patera:1976ud}, and leave a 
more complete analysis for a future work. 

A note on notation is in order here. The Lie algebras under 
study will be denoted as ${\cal A}_{d,i}$, where $d$ is the dimension of the algebra (the number of its generators) 
and $i$ is just an enumerative index according to the tables of \cite{Patera:1976ud}. Moreover, when there is some parameter on which the algebra depends, 
it will appear as superscript, e.g. ${\cal A}_{d,i}^{\a}$ if there is one parameter $\a$. Let us also note that the 
generators of an algebra will be denoted as $X_a,a=1,\dots,d$.

For the solutions that we find, the corresponding quadratic Casimir operators are presented, as well as 
the Killing form $g_{ab}$ and the invariant metric $\Omega_{ab}$ (whenever it exists). The related definitions appear in the appendix. It is important to note that unlike semisimple Lie groups, where 
the invariant metric is proportional to the Killing form, for a general Lie group this is not true as it will become 
obvious in some of the following examples. 

As a final remark, let us explain that when we refer to a solution of the equations (\ref{ikkteoms}) it is implied that the rest of the matrices (i.e. ten minus the number of algebra generators) are taken 
to be zero. For example, in the case of a 3-dimensional algebra with generators $X_1,X_2$ and $X_3$, 
a solution would be a set of commutation relations which solve eq. (\ref{ikkteoms}) accompanied by $X_i=0,~i=0,4,\dots,9$. Of course such solutions may be subsequently combined with the basic solution of 
section 2.

%\subsection{Solutions based on Lie algebras of dimension up to six}
\subsection{Solutions of the equations of motion}

\paragraph{1- and 2-dimensional Lie algebras.}

There is one real 1-dimensional Lie algebra, ${\cal A}_{1,1}$. Evidently, this is an abelian algebra and it 
constitutes a solution of the IKKT model, albeit not an interesting one as we have already argued. Therefore it will 
not be considered further.

As far as 2-dimensional Lie algebras are concerned, there exists the algebraic sum of two copies of ${\cal A}_{1,1}$, 
namely ${\cal A}_{1,1}\oplus {\cal A}_{1,1}$, which we shall not consider for the above reasons. This will be true in all dimensions
 to follow from now on and algebraic sums of lower-dimensional abelian algebras will not be considered further.
Moreover, there exists one non-abelian solvable Lie algebra in two dimensions, based on the 
following commutation relation,
\be 
[X_1,X_2]=iX_2.
\ee 
However, it is clearly not a solution to the matrix model, since one may easily verify that 
\be 
[X_1,[X_1,X_2]]=-X_2\ne 0
\ee
and therefore the corresponding equation of motion is not satisfied.

\paragraph{3-dimensional Lie algebras.}

There exist nine real Lie algebras ${\cal A}_{3,i}$ in three dimensions\footnote{We always refer to algebras which are not algebraic sums of lower-dimensional ones.}. One of them is nilpotent, six are solvable and two are the well-known semisimple ones, which are isomorphic to $su(2)$ and $su(1,1)\sim sl(2;\R)$. 
Using their commutation relations appearing in \cite{Patera:1976ud} it is straightforward to verify that 
only one of them, the ${\cal A}_{3,1}$ one, provides a solution of the equations of motion (\ref{ikkteoms}). The only 
nontrivial commutation relation of this algebra is 
\be [X_2,X_3]=iX_1. \ee 
This algebra is nilpotent and its only quadratic Casimir operator is $C^{(2)}({\cal A}_{3,1})=X_1^2$. Its Killing form vanishes identically, $g_{ab}=0$, while 
\be \Omega^{ab}=\mbox{diag}(1,0,0), \ee 
which is degenerate.

\paragraph{4-dimensional Lie algebras.}

There are twelve 4-dimensional real Lie algebras ${\cal A}_{4,i}$ and in particular one nilpotent and eleven solvable ones. Out of these algebras we find only one that provides a solution to the equations (\ref{ikkteoms}). It 
is the ${\cal A}_{4,12}$ one, which is solvable and its commutation relations are
\be 
[X_1,X_3]=iX_1, \quad [X_2,X_3]=iX_2, \quad [X_1,X_4]=-iX_2, \quad [X_2,X_4]=iX_1.
\ee
However, this algebra does not possess any quadratic Casimir operators (in fact it does not possess any 
invariants at all) and therefore it is of no further interest.

\paragraph{5-dimensional Lie algebras.}

In five dimensions the number of real Lie algebras sums up to forty. Six of them are nilpotent and the 
rest are solvable. Scanning the commutation relations of the corresponding table in \cite{Patera:1976ud} 
we find three solutions to the equations (\ref{ikkteoms}), corresponding to the algebras ${\cal A}_{5,1},{\cal A}_{5,4}$ 
and ${\cal A}_{5,39}$.

${\cal A}_{5,1}$ is a nilpotent algebra with the following commutation relations,
\be 
[X_3,X_5]=iX_1, \quad [X_4,X_5]=iX_2.
\ee
Its invariants are $X_1,X_2$ and $X_2X_3-X_1X_4$ and therefore its quadratic Casimir operator may be written as
\be C^{(2)}({\cal A}_{5,1})=pX_1^2+qX_2^2+r(X_2X_3-X_1X_4), \ee
where $p,q,r$ are arbitrary real parameters. Since the algebra is nilpotent its Killing form is identically zero, 
while it holds that
\be
\Omega^{ab}=\begin{pmatrix}
       p & 0 & 0 & -r & 0 \\ 0 & q & r & 0 & 0 \\ 0 & -r & 0 & 0 & 0 \\ r & 0 & 0 & 0 & 0 \\ 0 & 0 & 0 & 0 & 0 
      \end{pmatrix},
\ee
which is again degenerate and does not possess an inverse.

For ${\cal A}_{5,4}$, which is also nilpotent, the commutation relations are
\be [X_2,X_4]=iX_1, \quad [X_3,X_5]=iX_1. \ee
Its only invariant is $X_1$ and therefore it holds that 
\be C^{(2)}({\cal A}_{5,4})=X_1^2, \ee
which leads to 
%$g_{ab}=0$ and 
$\Omega^{ab}=\mbox{diag}(1,0,0,0,0)$.

Finally, ${\cal A}_{5,39}$ is solvable with commutation relations
\be [X_1,X_4]=iX_1, \quad [X_2,X_4]=iX_2, \quad [X_1,X_5]=-iX_2, \quad [X_2,X_5]=iX_1, \quad [X_4,X_5]=iX_3. \ee
Its only invariant is $X_3$ and therefore we find $C^{(2)}({\cal A}_{5,39})=X_3^2$, $g_{ab}=\mbox{diag}(0,0,0,2,-2)$ 
and $\Omega^{ab}=\mbox{diag}(0,0,1,0,0)$. Tha last two cases are obviously degenerate too.

\paragraph{6-dimensional nilpotent Lie algebras.}

There are twenty-two real nilpotent Lie algebras of dimension six which are not algebraic sums of 
lower-dimensional ones. Four of them provide solutions to the 
equations (\ref{ikkteoms}) and in particular the ${\cal A}_{6,3},{\cal A}_{6,4},{\cal A}_{6,5}^{\a}$ and ${\cal A}_{6,14}^{-1}$. In the two latter 
cases the algebras have a continuous parameter $\a$ which in the last case is fixed to $-1$ in order to provide the desired 
solution. Let us mention again that the Killing form for all the nilpotent Lie algebras is identically zero.

The algebra ${\cal A}_{6,4}$ has the following commutation relations,
\be 
[X_1,X_2]=iX_5, \quad [X_1,X_3]=iX_6, \quad [X_2,X_4]=iX_6.
\ee
Its invariants are $X_5$ and $X_6$ and therefore 
\be C^{(2)}({\cal A}_{6,4})=pX_5^2+qX_6^2, \ee
which gives the degenerate metric
\be \Omega^{ab}=\mbox{diag}(0,0,0,0,p,q). \ee

Similarly, for ${\cal A}_{6,5}^{\a}$ the commutation relations read as
\be 
[X_1,X_3]=iX_5, \quad [X_1,X_4]=iX_6, \quad [X_2,X_3]=i\a X_6, \quad [X_2,X_4]=iX_5, \quad \a\ne 0.
\ee
The invariants are again $X_5$ and $X_6$ and therefore the results of the previous case apply in the present 
one as well.

The algebra ${\cal A}_{6,14}^{-1}$ has the following commutation relations, 
\be 
[X_1,X_3]=iX_4, \quad [X_1,X_4]=iX_6, \quad [X_2,X_3]=iX_5, \quad [X_2,X_5]=-iX_6,
\ee
with invariants $X_6$ and $X_5^2-X_4^2+2X_3X_6$. Therefore the quadratic Casimir operator is
\be
C^{(2)}({\cal A}_{5,14}^{-1})=pX_6^2+q(X_5^2-X_4^2+2X_3X_6), \ee
which gives
\be 
\Omega^{ab}=\begin{pmatrix}
       0 & 0 & 0 & 0 & 0 & 0 \\ 0 & 0 & 0 & 0 & 0 & 0 \\ 0 & 0 & 0 & 0 & 0 & q
	\\ 0 & 0 & 0 & -q & 0 & 0 \\ 0 & 0 & 0 & 0 & q & 0 \\ 0 & 0 & -q & 0 & 0 & p 
      \end{pmatrix},
\ee
which is again degenerate.

The most interesting solution corresponds to the algebra ${\cal A}_{6,3}$. This one has the following commutation 
relations,
\be 
[X_1,X_2]=iX_6, \quad [X_1,X_3]=iX_4, \quad [X_2,X_3]=iX_5.
\ee
Its invariants are $X_4,X_5,X_6$ and $X_1X_5+X_3X_6-X_2X_4$ and therefore the general form of its quadratic 
Casimir operator reads as
\be C^{(2)}({\cal A}_{6,3})=pX_4^2+qX_5^2+rX_6^2+s(X_1X_5+X_3X_6-X_2X_4). \ee
Now the corresponding metric is given by
\be 
\Omega^{ab}=\begin{pmatrix}
       0 & 0 & 0 & 0 & s & 0 \\ 0 & 0 & 0 & -s & 0 & 0 \\ 0 & 0 & 0 & 0 & 0 & s
	\\ 0 & -s & 0 & p & 0 & 0 \\ s & 0 & 0 & 0 & q & 0 \\ 0 & 0 & s & 0 & 0 & r 
      \end{pmatrix},
\ee
which is the first non-degenerate case that we encounter in our analysis. The determinant of the metric 
is 
\be |\Omega^{ab}|=-s^6 \ne 0 \ee
and therefore it is invertible with inverse
\be 
\Omega_{ab}=\begin{pmatrix}
       -q/s^2 & 0 & 0 & 0 & 1/s & 0 \\ 0 & -p/s^2 & 0 & -1/s & 0 & 0 \\ 0 & 0 & -r/s^2 & 0 & 0 & 1/s
	\\ 0 & -1/s & 0 & 0 & 0 & 0 \\ 1/s & 0 & 0 & 0 & 0 & 0 \\ 0 & 0 & 1/s & 0 & 0 & 0 
      \end{pmatrix}.
\ee
The six eigenvalues of the latter are
\be 
\frac 1{s^2}\biggl(-x_i\pm\sqrt{x_i^2+4s^2}\biggl), \quad i=1,2,3,
\ee
where $x_1=p$, $x_2=q$ and $x_3=r$. We observe that there are three positive and three negative eigenvalues, 
therefore the algebra is non-compact. 
%This algebra will be discussed further in the following subsection, 
%where a non-commutative space based on it will be constructed.

Finally, let us close this subsection by mentioning that there is one further non-trivial case in six dimensions, 
which is the algebraic sum of two 3-dimensional nilpotent algebras ${\cal A}_{3,1}$, namely ${\cal A}_{3,1}\oplus {\cal A}_{3,1}$. The properties of this case are directly derived from the properties of 
${\cal A}_{3,1}$, which were presented before.

\subsection{Representations in terms of Hermitian matrices}

Let us now discuss the representation of the above algebras in terms of Hermitian matrices. First let us note 
that a complete study of the unitary representations of nilpotent Lie groups was performed in \cite{kirillov}, 
which facilitates our task.

The method that will be followed consists of the following steps. First the central elements, i.e. the elements
which commute with all the algebra generators, are identified. These elements are mapped to operators which are 
multiples of the identity. Then, in order to completely define the representation one has to map the remaining 
elements to Hermitian matrices. It turns out that this last step amounts in mapping these elements to the usual 
operators for coordinates and momenta in quantum mechanics. Let us now present a case by case analysis 
following the above steps for the algebras which were identified in the previous subsection.

\paragraph{${\cal A}_{3,1}$ case.}  This algebra has one central element, the $X_1$. Therefore we map this 
element to a multiple of the identity, 
\be X_1=\theta\one, \quad \theta \in \R. \ee
The remaining elements now satisfy the commutation relation, 
\be\label{ccr} [X_2,X_3]=i\theta\one, \ee
which reduces to the Moyal-Weyl case. Clearly, $X_2$ and $X_3$ may then be represented by the usual Hermitian 
matrices corresponding to the coordinate and momentum operators of quantum mechanics. These matrices are of 
course infinite-dimensional and they have the well-known form
\be P=\sqrt{\frac{1}{ 2}}\begin{pmatrix} 0&1&0&0&0&\dots \\ 
1&0&\sqrt 2&0&0&\dots \\
0&\sqrt 2&0&\sqrt 3&0&\dots \\ 0&0&\sqrt 3&0&0&\dots \\
\vdots \end{pmatrix}, 
\quad Q=\sqrt{\frac{1}2}\begin{pmatrix} 0&i&0&0&0&\dots \\ 
-i&0&i\sqrt 2&0&0&\dots \\
0&-i\sqrt 2&0&i\sqrt 3&0&\dots \\ 0&0&-i\sqrt 3&0&0&\dots \\
\vdots\end{pmatrix}. \ee 
Then the solution we have obtained is
\be 
\{X_1=\theta\one, \quad X_2=\sqrt\theta Q, \quad X_3=\sqrt\theta P\}. \ee

%In the following it will turn out that in all the tractable cases canonical commutation relations of the 
%form (\ref{ccr}) will have to be satisfied. The representation in terms of Hermitian matrices is then 
%obvious. This is nothing but a manifestation of the Stone-von Neumann theorem \cite{Stone,vonneumann}.

\paragraph{${\cal A}_{4,12}$ case.} As we already mentioned in the previous subsection, this algebra does not 
possess any invariants and therefore the method we follow here cannot be applied. 

\paragraph{${\cal A}_{5,1}$ case.} The central elements of this algebra are $X_1$ and $X_2$ and the combination 
$X_2X_3-X_1X_4$. Therefore we set
\be X_1=\theta_1\one \quad \mbox{and} \quad X_2=\theta_2\one. \ee
Then the commutation relations read as
\be [X_3,X_5]=i\theta_1\one, \quad [X_4,X_5]=i\theta_2\one. \ee 
Moreover the last quadratic invariant has to be fixed,
\be X_2X_3-X_1X_4=\theta\one. \ee 
The resulting solution is 
\be 
\{X_1=\theta_1\one, \quad X_2=\theta_2\one, \quad X_3=\sqrt{\theta_1}Q, \quad X_4=\frac{\theta_2}{\sqrt{\theta_1}}Q-\frac{\theta}{\theta_1}\one, \quad X_5=\sqrt{\theta_1}P\}.
\ee

\paragraph{${\cal A}_{5,4}$ case.} 

The unique central element in the present case is $X_1$. Therefore we set
\be X_1=\theta\one, \ee 
which leads to the commutation relations
\be [X_2,X_4]=i\theta\one, \quad [X_3,X_5]=i\theta\one. \ee
These relations may be interpreted as two quantum planes in the directions $(24)$ and $(35)$ respectively, 
with the same quantization parameter $\theta$. The solution is
\be 
\{X_1=\theta\one, \quad X_2=\sqrt{\theta}Q, \quad X_3=\sqrt{\theta}Q', \quad X_4=\sqrt\theta P, \quad X_5=\sqrt\theta P'\},
\ee
where $(Q,P)$ and $(Q',P')$ are two sets of Hermitian matrices, representing the two different quantum planes and therefore mutually commuting.

\paragraph{${\cal A}_{5,39}$ case.} This is a solvable Lie algebra and the method we follow here does not 
directly apply. Therefore this case is less clear and it will not be treated any further.

\paragraph{${\cal A}_{6,3}$ case.} The central elements in the present case are $X_4,X_5,X_6$ and $X_1X_5+X_3X_6-X_2X_4$. Therefore we set
\be X_4=\theta_4\one, \quad X_5=\theta_5\one \quad \mbox{and} \quad X_6=\theta_6\one. \ee 
The commutation relations take the form
\be [X_1,X_2]=i\theta_6\one, \quad [X_1,X_3]=i\theta_4\one, \quad [X_2,X_3]=i\theta_5\one, \ee 
while the quadratic invariant is fixed according to
\be X_1X_5+X_3X_6-X_2X_4=\theta\one. \ee
The resulting solution in this case is
\bea
&&\{ X_1=\sqrt{\theta_6}Q, \quad X_2=\sqrt{\theta_6}P, \quad X_3=-\frac{\theta_5}{\sqrt{\theta_6}}Q
+\frac{\theta_4}{\sqrt{\theta_6}}P+\frac{\theta}{\theta_6}\one, \nn\\ 
&& X_4=\theta_4\one, \quad X_5=\theta_5\one, \quad X_6=\theta_6\one
\}.
\eea

\paragraph{${\cal A}_{6,4}$ and ${\cal A}_{6,5}^{\a}$ cases.} The central elements for these algebras are $X_5$ and $X_6$. Therefore in both cases we set
\be X_5=\theta_5\one, \quad X_6=\theta_6\one. \ee 
Then, for the first case we obtain the commuation relations 
\be [X_1,X_2]=i\theta_5\one, \quad [X_1,X_3]=i\theta_6\one, \quad [X_2,X_4]=i\theta_6\one, \ee 
while for the second case
\be [X_1,X_3]=i\theta_5\one, \quad [X_1,X_4]=i\theta_6\one, \quad [X_2,X_3]=i\a\theta_6\one, \quad [X_2,X_4]=i\theta_5\one. \ee 
The resulting solutions take the following form,
\bea 
&&\{X_1=\sqrt{\theta_6}Q-\frac{\theta_5}{2\sqrt{\theta_6}}P', \quad X_2=\sqrt{\theta_6}Q'+\frac{\theta_5}{2\sqrt{\theta_6}}P, \nn\\ 
&& X_3=\sqrt{\theta_6}P, \quad X_4=\sqrt{\theta_6}P', \quad X_5=\theta_5\one, \quad X_6=\theta_6\one\},
\eea
and 
\bea 
&&\{X_1=\sqrt{\theta_5}Q+\frac{\theta_6}{\sqrt{\theta_5}}Q', \quad X_2=\sqrt{\theta_5}Q'+\frac{\a\theta_6}{\sqrt{\theta_5}}Q, \nn\\ 
&& X_3=\sqrt{\theta_5}P, \quad X_4=\sqrt{\theta_5}P', \quad X_5=\theta_5\one, \quad X_6=\theta_6\one\},
\eea
respectively, where $(Q,P)$ and $(Q',P')$ are again two sets of mutually commuting representations.

\paragraph{${\cal A}_{6,14}^{-1}$ case.} The central elements in this case are $X_6$ and $X_5^2-X_4^2+2X_3X_6$ and we set 
\be X_6=\theta_6\one. \ee
Then the commutation relations read as 
\be [X_1,X_3]=iX_4, \quad [X_1,X_4]=i\theta_6\one, \quad [X_2,X_3]=iX_5, \quad [X_2,X_5]=-i\theta_6\one, \ee 
 while there is also a relation of the form
\be X_5^2-X_4^2+2X_3X_6=\theta\one. \ee
The solution in this case is 
\bea 
&&\{X_1=\sqrt{\theta_6}Q, \quad X_2=\sqrt{\theta_6}P', 
\quad X_3=\frac 12 (P^2-Q'^2+\frac{\theta}{\theta_6}\one),\nn\\ && X_4=\sqrt{\theta_6}P, \quad X_5=\sqrt{\theta_6}Q', \quad X_6=\theta_6\one\}.
\eea

%In the following subsection we shall discuss how the above relations are associated to NC spaces. 

Having identified the above Lie-algebraic solutions of the IKKT model, in the following subsection we 
discuss their relation to NC geometry in a more general context.

\subsection{Non-commutative spaces and $\ast$-products}

\paragraph{NC spaces.}

The construction of ``non-commutative (NC) spaces'' is based on a shift from the space itself to the 
algebra of functions defined on it \cite{Connes:1994yd,Madore:2000aq,Szabo:2001kg,Landi:1997sh,Balachandran:2005ew}. 
Therefore, strictly speaking a NC space is not a space in the classical sense but instead it corresponds 
to an associative but not necessarily commutative algebra $\cal A$, accompanied with a set of relations.

In order to be more specific, let us consider an associative algebra $\cal A$ with generators $X_a,a=1,\dots,N$. 
These generators satisfy certain commutation relations of the general form
\be [X_a,X_b]=i\theta_{ab}(X), \ee
where $\theta_{ab}(X)$ is an arbitrary function of the generators $X_a$. Then the above algebraic structure defines a 
NC space and the generators of the algebra are commonly referred to as ``coordinates'' of the NC space \cite{Madore:2000en}. The case of constant $\theta_{ab}$ corresponds to the Moyal-Weyl quantum plane, 
which was encountered in section 2 as the basic solution of the IKKT MM.
%Let us make here the following important remark. In order for the NC space to be 
%well-defined one has to fix the representation of the algebra as well. Usually this is done by fixing 
%the corresponding Casimir operator of the algebra. We shall return to this point below.

The cases we already studied in the present section correspond to a Lie algebra structure. 
In other words, the function $\theta_{ab}$ is linear in the generators $X_a$ and the commutation relations 
read as
\be \label{liealgstr}
[X_a,X_b]=if_{ab}^{ \ \ c}X_c. \ee
The most prominent representatives of such a structure are the fuzzy 2-sphere \cite{Madore:1991bw} and its higher-dimensional generalizations \cite{Ramgoolam:2001zx},
 the fuzzy complex projective spaces \cite{Grosse:1999ci,Balachandran:2001dd} and others \cite{Trivedi:2000mq,Dolan:2001mi,Murray:2006pi}. All these NC spaces are compact, since 
they are based on compact semisimple Lie algebras. As we saw, these compact NC spaces do not directly provide solutions of the undeformed IKKT 
MM\footnote{See however \cite{Steinacker:2011wb}.}. However, we proved that there exist solutions of the IKKT 
model with the structure (\ref{liealgstr}), albeit based on non-compact algebras. Indeed, it is obvious 
from our previous analysis that in each case there is a set of generators and relations, along with the 
prescribed Casimir operators which fix the representation of the algebra. Thus all the cases that were discussed 
correspond to well-defined NC spaces. Moreover, 
in all the cases the number of generators minus the number of invariants (whose fixing specifies the representation) is always even. This means that the resulting solutions describe non-compact, non-commutative even-dimensional branes, similarly to the $Dp$-brane solutions of the IIB string theory (with $p$ odd), appropriately embedded in 10-dimensional $\R^{10}$.
%The next step is to ask 
%whether NC spaces based on these algebras can be constructed.
%, especially compact ones. In the following 
%this question will be answered in the affirmative.
%The next step is to 
%construct, whenever it is possible, NC spaces based on these algebras, which we do in the following paragraph. 

\paragraph{Weyl quantization and $\ast$-products.}

As we discussed above, the shift from spaces to algebras paves the road to NC geometry and provides a natural 
set-up to construct NC/quantized spaces which correspond to certain algebraic structures. 
%The same set-up may account 
%on the problem of the quantization of a classical manifold. 
Indeed, a natural way to quantize a manifold 
is to consider an appropriate algebra of functions on it and instead quantize the algebra, either by 
truncating it or deforming its product structure. The latter possibility belongs in the broad context of 
deformation quantization \cite{Bayen:1977ha}, whose most prominent physical example is phase-space (Weyl) 
quantization \cite{Weyl:1927vd}. 

Let us briefly discuss Weyl quantization in the case of a Lie algebra structure following \cite{Madore:2000en} and apply it in the specific cases studied here. A more 
formal and rigorous discussion based on the pioneering work of Kontsevich \cite{Kontsevich:1997vb} may be found in \cite{kathotia}.
Let us 
denote classical (commutative) coordinates by $x^a,a=1,\dots,N$ and elements of $\cal A$ (NC coordinates) 
by $X^a$, as before. An operator $W(f)$ may be associated to every classical function $f(x)$, given by
\be \label{weylop}
W(f)=\frac{1}{\sqrt{(2\pi)^n}}\int d^nke^{ik_aX^a}\tilde f(k),
\ee 
where $\tilde f$ is the Fourier transform of $f$,
\be 
\tilde f(k)=\frac{1}{\sqrt{(2\pi)^n}}\int d^nxe^{ik_ax^a}f(x).
\ee
Multiplying operators of the kind appearing in (\ref{weylop}) results in new operators, which might or might 
not be associated to classical functions as well. In the case that this is possible, i.e. when 
\be W(f)W(g)=W(h), \ee 
the corresponding function $h$ will be identified with a deformed product of $f$ and $g$, which is denoted by 
$\ast$, i.e.
\be h=f\ast g. \ee
More explicitly, we can write the product of the operators as
\be
W(f)W(g)=\frac{1}{(2\pi)^n}\int d^nkd^npe^{ik_aX^a}e^{ip_bX^b}\tilde f(k)\tilde g(p).
\ee
Then the function $f\ast g$ exists if the product of the two exponentials in the integrand can be calculated
by the Baker-Campbell-Hausdorff formula. In the case of a Lie structure one can write
\be 
e^{ik_aX^a}e^{ip_bX^b}=e^{iP_a(k,p)X^a},
\ee
where 
\be P_a=k_a+p_a+\frac 12 g_a(k,p), \ee
where the $g_a$ contain the information about the NC structure. Having determined the functions $g_a$ it is 
straightforward to write down the explicit formula for the $\ast$-product,
\be\label{liestar} f\ast g=e^{\frac{i}{2}x^ag_a(i\frac{\partial}{\partial y},i\frac{\partial}{\partial z})}f(y)g(z)|_{y,z\rightarrow x}. \ee
%Later in this section we shall determine the functions $g_a$ for a class of manifolds based on the Lie 
%algebras which we met in section 3. 

%\paragraph{$\ast$-products.}
%Having proven the existence of compact nilmanifolds based on the nilpotent Lie algebras which solve the 
%equations of motion of the IKKT MM, let us discuss here their quantization\footnote{For the example based 
%on ${\cal A}_{3,1}$ see \cite{Rieffel}.}. Following our previous 
%discussion, we shall consider the algebra of classical functions on the nilmanifold and deform its 
%product structure by providing a $\ast$-product. As it was shown in section 4.1, this boils down to 
%the determination of the functions $g_a(k,p)$.

Let us now determine the functions $g_a$ for the cases that are studied here.
By direct calculation  we obtain the following results for each of the seven nilpotent cases:
\begin{description}\itemsep10pt
\item[${\cal A}_{3,1}:$] $g_1=ik_{[2}p_{3]}$, \quad $g_{2,3}=0.$
\item[${\cal A}_{5,1}:$] $g_1=ik_{[3}p_{5]}$, \quad $g_2=ik_{[4}p_{5]}$, \quad $g_{3,4,5}=0.$
\item[${\cal A}_{5,4}:$] $g_1=i(k_{[2}p_{4]}+k_{[3}p_{5]})$, \quad $g_{2,3,4,5}=0.$
\item[${\cal A}_{6,3}:$] $g_{1,2,3}=0$, \quad $g_4=ik_{[1}p_{3]}$, \quad $g_5=ik_{[2}p_{3]}$, \quad $g_6=ik_{[1}p_{2]}$.
\item[${\cal A}_{6,4}:$] $g_{1,2,3,4}=0$, \quad $g_5=ik_{[1}p_{2]}$, \quad $g_6=i(k_{[1}p_{3]}+k_{[2}p_{4]})$.
\item[${\cal A}_{6,5}:$] $g_{1,2,3,4}=0$, \quad $g_5=i(k_{[1}p_{3]}+k_{[2}p_{4]})$, \quad $g_6=i(k_{[1}p_{4]}+k_{[2}p_{3]})$.
\item[${\cal A}_{6,14}:$] $g_{1,2,3}=0$, \quad $g_4=ik_{[1}p_{3]}$, \quad $g_5=ik_{[2}p_{3]}$, \\ \qquad $g_6=i\biggl(k_{[1}p_{4]}-k_{[2}p_{5]}+\frac i6\big(k_1k_{[1}p_{3]}-k_2k_{[2}p_{3]}-p_1k_{[1}p_{3]}+p_2k_{[2}p_{3]}\big)\biggl)$,
\end{description}
where the brackets appearing in the subscripts denote antisymmetrization with weight one.
Plugging these expressions in the equation (\ref{liestar}) gives directly the corresponding $\ast$-product 
in each case. It is worth noting that due to the nature of nilpotent algebras the Baker-Campbell-Hausdorff 
formula terminates and the above functions determine exactly the exponent of the $\ast$-product.

\section{Towards quantized compact nilmanifolds}

% In the previous section we identified nine algebras (plus one case corresponding to the algebraic sum of two 
% of them) which constitute classical solutions to the equations of 
% motion of the IKKT MM. The next step is to study whether it is possible to construct non-commutative (NC) 
% spaces based on these algebras. 
% %Since these spaces will correspond to NC brane solutions of the MM, we are interested in the 
% %cases with even number of dimensions, i.e. four and six. This leaves us with five cases, one 4-dimensional 
% %and four 6-dimensional. Moreover, 
% We shall focus our interest on the construction of compact NC spaces. Possible solutions based on compact 
% spaces may subsequently be combined with the basic solution which was described in section 2 in order to 
% form new solutions of the MM with compactified extra dimensions.
% %In the following, first a brief general description of NC spaces 
% %and Weyl quantization is given, as well as some discussion on nilmanifolds and solvmanifolds. 
% %Subsequently, a case by case analysis on the quantization of the symplectic structure of the nilmanifolds 
% %based on the Lie algebras which constitute solutions of the MM is performed. 

%\subsection{Nilmanifolds and solvmanifolds}
In the present section we deviate from the search for solutions of the IKKT model and we pose the following 
question: are there compact manifolds based on the algebras discussed in section 3?

The simplest way to construct a manifold out of a nilpotent or a solvable Lie algebra $\cal A$ is to consider the 
action of a dicrete (sub)group $\G$ on its Lie group. Then the quotient\footnote{Here $A$ denotes the group 
associated to the algebra $\cal A$.} $M=A/\G$ is called a nilmanifold or a 
solvmanifold respectively. 
A very important result for our purposes states that in the nilpotent case such a 
construction is possible if and only if the corresponding Lie algebra has rational structure constants in 
some basis \cite{malcev}. This guarantees that in the nilpotent cases that we study here, a $\G$ as above always 
exists\footnote{This does not mean though that it has to be unique.}. For the solvable cases the situation is 
more complicated but it will soon become evident that the ones we met in section 3 are not of further interest 
for our purposes.

An important issue which we would like to mention regards the compactness of a nilmanifold. It is true that 
even starting with a non-compact group $A$ it is possible to construct a compact manifold by considering 
its quotient by a compact discrete subgroup of it. A necessary condition for compactness is that the 
group is unimodular, i.e. its structure constants satisfy $f^a_{\ ab}=0$ (this was already discussed in
 \cite{Scherk:1979zr}). This condition is not sufficient 
but for nilpotent groups the requirement of rational structure constants is enough. 

As a first check on whether we can construct compact manifolds based on the algebras that were singled out in 
section 3, 
%as solutions to the equations (\ref{ikkteoms}) 
let us try to verify the condition of unimodularity. It is 
straightforward to check (by mere inspection of the commutation relations) that ${\cal A}_{3,1},{\cal A}_{5,1},{\cal A}_{5,4},{\cal A}_{6,3},{\cal A}_{6,4},{\cal A}_{6,5}^{\a}$ and ${\cal A}_{6,14}^{-1}$ indeed pass the 
test, while on the other hand ${\cal A}_{4,12}$ and ${\cal A}_{5,39}$ fail to do so. Therefore the two latter 
cases do not give rise to compact manifolds. It is worth noting that these two cases are exactly the only 
solvable ones that we found in section 3 and therefore our present analysis shows that there are no compact 
solvmanifolds\footnote{This argument holds of course only for up to 5-dimensional solvable algebras which we consider here.}  corresponding to algebras which solve the equations (\ref{ikkteoms}). Therefore in the following 
only nilmanifolds will be discussed. For tables of nilmanifolds and solvmanifolds in six dimensions the reader 
may consult \cite{gualtieri,Grana:2006kf}.

Let us proceed by giving two explicit examples of the construction of a nilmanifold. The first one corresponds to the 
simplest case of the algebra ${\cal A}_{3,1}$ with non-trivial commutation relation $[X_2,X_3]=X_1$ \cite{Rieffel}. A basis 
for the algebra is given by the following $3\times 3$ upper triangular matrices\footnote{Of course these matrices are not Hermitian and therefore they cannot be related to solutions of the IKKT model.},
\be
X_1=\begin{pmatrix} 0 &0&1 \\ 0&0&0 \\0&0&0 \end{pmatrix}, \quad 
X_2=\begin{pmatrix} 0 &1&0 \\ 0&0&0 \\0&0&0 \end{pmatrix}, \quad 
X_3=\begin{pmatrix} 0 &0&0 \\ 0&0&1 \\0&0&0 \end{pmatrix}.
 \ee
Then, any element of the corresponding group $A_{3,1}$ may be parametrized as 
\be \label{groupelement}
g=\begin{pmatrix} 1 &x_2&x_1 \\ 0&1&x_3 \\0&0&1 \end{pmatrix}.
\ee
This is clearly a non-compact group. According to the above discussion, in order to produce a compact manifold 
out of it, a compact discrete subgroup $\G$ has to be considered. Such a subgroup is given by those elements $g\in A_{3,1}$ which have integer values of $x_2,x_3$ and $c x_1$, where $c$ is a positive integer. 
%In other words, 
%the group $A_{3,1}$ is compactified by making the following identifications,
%\be 
%(a,b,c)=(a+i_1,b,c)=(a,b,c)=
%\ee
 Then the 
quotient $A_{3,1}/\G$ is indeed a compact nilmanifold. In the physics literature this manifold is known as 
a twisted torus and it corresponds to a (twisted) fibration of a torus over another torus, see e.g. \cite{Grana:2006kf}. 

Let us explain the above construction in more detail. Consider a representative element $g\in A_{3,1}$, 
as in (\ref{groupelement}). Then 
the Maurer-Cartan 1-form $e$ is given by
\be 
e=g^{-1}dg,
\ee
which gives
\be 
e=\begin{pmatrix} 0&dx_2&dx_1-x_2dx_3 \\ 0&0&dx_3 \\ 0&0&0 \end{pmatrix}. \ee
The 1-form $e$ is Lie-algebra valued, $e=e^aX_a$, and its components are 
\be e^1=dx^1-x_2dx_3, \quad e^2=dx_2, \quad e^3=dx_3, 
\ee
which evidently satisfy the Maurer-Cartan equations 
\be de^a=-\frac 12 f_{bc}^ae^b\wedge e^c, \ee since $de^2=de^3=0$ and $de^1=-e^2\wedge e^3$.
The important observation here is that in order to compactify the group one has to introduce a twist. Indeed, 
while for the directions $x_1$ and $x_3$ the compactification is achieved by the identifications 
\be\label{id1} (x_1,x_2,x_3)\sim (x_1+a,x_2,x_3)\sim(x_1,x_2,x_3+b), \quad a,b\in \Z,\ee 
one cannot do the same for $x_2$, i.e. the identification $(x_1,x_2,x_3)\sim(x_1,x_2+c,x_3)$ obviously does 
not work. Instead, the correct identification is
\be\label{id2} (x_1,x_2,x_3)\sim(x_1+cx_3,x_2+c,x_3). \ee
Under (\ref{id1}) and (\ref{id2}) the desired (twisted) compactification is achieved.

The above example serves as a prototype for any other. One can always write down a basis for the algebra 
in terms of upper triangular matrices and compactify the corresponding group by modding out a discrete subgroup 
corresponding to elements with integer entries. Let us work out in some detail a less trivial, 6-dimensional 
example, based on the algebra ${\cal A}_{6,3}$. A basis for this algebra is given by the following $6\times 6$ 
upper triangular matrices:
\bea
X_1&=&\begin{pmatrix} 0&0&1&0&0&0 \\ &0&0&0&0&0 \\ &&0&0&0&0 \\ &&&0&0&0 \\ &&&&0&0 \\&&&&&0 \end{pmatrix},
\quad 
X_2=\begin{pmatrix} 0&1&0&0&0&0 \\ &0&0&0&0&0 \\ &&0&1&0&0 \\ &&&0&0&0 \\ &&&&0&0 \\&&&&&0 \end{pmatrix},
\quad 
X_3=\begin{pmatrix} 0&0&0&0&0&0 \\ &0&0&0&1&0 \\ &&0&0&0&1 \\ &&&0&0&0 \\ &&&&0&0 \\&&&&&0 \end{pmatrix},
\nn \\
X_4&=&\begin{pmatrix} 0&0&0&0&0&1 \\ &0&0&0&0&0 \\ &&0&0&0&0 \\ &&&0&0&0 \\ &&&&0&0 \\&&&&&0 \end{pmatrix},
\quad
X_5=\begin{pmatrix} 0&0&0&0&1&0 \\ &0&0&0&0&0 \\ &&0&0&0&0 \\ &&&0&0&0 \\ &&&&0&0 \\&&&&&0 \end{pmatrix},
\quad 
X_6=\begin{pmatrix} 0&0&0&1&0&0 \\ &0&0&0&0&0 \\ &&0&0&0&0 \\ &&&0&0&0 \\ &&&&0&0 \\&&&&&0 \end{pmatrix}.\nn \\
\eea
The corresponding general group element is given by
\be
g=\begin{pmatrix} 1&x_2&x_1&x_6&x_5&x_4 \\ &1&0&0&x_3&0 \\ &&1&x_2&0&x_3 \\ &&&1&0&0 \\ &&&&1&0 \\&&&&&1 \end{pmatrix}.
\ee
The Maurer-Cartan 1-form may be computed and it has the following form,
\be
e=\begin{pmatrix} 0&dx_2&dx_1&dx_6-x_1dx_2&dx_5-x_2dx_3&dx_4-x_1dx_3 \\ &0&0&0&dx_3&0 \\ &&0&dx_2&0&dx_3 \\ &&&0&0&0 \\ &&&&0&0 \\&&&&&0 \end{pmatrix},
\ee
with components
\bea
e^1&=&dx_1, \quad e^2=dx_2, \quad e^3=dx_3, \nn\\ \quad e^4&=&dx_4-x_1dx_3, \quad e^5=dx_5-x_2dx_3, \quad e^6=dx_6-x_1dx_2.
\eea
The necessity for twists is again evident. Indeed, it is straightforward to see that for the directions 
$x_3,x_4,x_5$ and $x_6$ we can consider the identifications
\bea\label{id3} (x_1,x_2,x_3,x_4,x_5,x_6)&\sim&(x_1,x_2,x_3+c,x_4,x_5,x_6) \nn \\
&\sim&(x_1,x_2,x_3,x_4+d,x_5,x_6) \nn\\ &\sim&(x_1,x_2,x_3,x_4,x_5+e,x_6) \nn \\ 
&\sim&(x_1,x_2,x_3,x_4,x_5,x_6+f), \quad c,d,e,f\in\Z,  \eea  
while for the $x_1$ and $x_2$ ones the correct identifications are
\bea\label{id4}
(x_1,x_2,x_3,x_4,x_5,x_6)&\sim&(x_1+a,x_2,x_3,x_4+ax_3,x_5,x_6+ax_2)\nn\\
&\sim&(x_1,x_2+b,x_3,x_4,x_5+bx_3,x_6), \quad a,b\in\Z.
\eea 
Under (\ref{id3}) and (\ref{id4}) we obtain the desired twisted compactification.

Following the above procedure, nilmanifolds corresponding 
to the seven nilpotent Lie algebras which were singled out in section 3 may be constructed. The deformation quantization 
of the first one, based on ${\cal A}_{3,1}$, was performed in \cite{Rieffel}. A detailed discussion on the  quantization of the rest of the cases is beyond the scope of the present paper. However, the pursue of this 
task would be of interest e.g. for compactifications of the IKKT model. In \cite{Connes:1997cr} it was shown 
that compactifications on NC tori correspond to supergravity backgrounds with constant three-form flux. Then 
one could expect that similar compactifications on NC twisted tori could account for non-geometric flux vacua
 \cite{Shelton:2005cf,Wecht:2007wu}, incorporating the non-geometric fluxes in the geometry of the NC space. We hope to report on this in a future 
publication.

\section{Discussion and Conclusions}

In the present paper we performed a survey of Lie-algebraic solutions to the IKKT matrix model. Up to now 
it was known that manifolds with Lie-type non-commutativity are either solutions of deformed MM \cite{Iso:2001mg,Kimura:2001uk,Kitazawa:2002xj,Azuma:2004zq,Steinacker:2003sd} or else 
some split non-commutativity has to be introduced \cite{Steinacker:2011wb}. Moreover, the above compact 
solutions are all based on compact semisimple Lie algebras. Our investigation revealed the possibility 
of obtaining  
(non-compact) solutions to the undeformed IKKT model without deformations or additional requirements, which
are based on nilpotent and solvable Lie algebras.
%, which may be associated to certain compact NC spaces, at 
%least in the nilpotent cases. 

More specifically, scanning the classification tables of \cite{Patera:1976ud} we found seven nilpotent 
Lie algebras (one 3-dimensional, two 5-dimensional and four 6-dimensional ones) and two solvable ones (one 
4-dimensional and one 5-dimensional) which solve the equations of motion of the IKKT model. Subsequently, 
we discussed the representation of these algebras by Hermitian matrices. This is always possible for the nilpotent cases, thus proving that they indeed 
constitute solutions to the IKKT model. It is straightforward to combine these solutions with the basic 4-dimensional solution of the 
IKKT model, which was presented in section 2 and corresponds to NC space-time as a Moyal-Weyl quantum plane 
$\R_{\theta}^4$.

In addition, we addressed the problem of constructing compact NC spaces associated to these algebras. The simplest 
constructions of compact spaces based on non-semisimple Lie algebras are the so-called nilmanifolds and 
solvmanifolds, also known as twisted tori in the physics literature. These correspond to quotients of the group of a nilpotent or solvable Lie algebra respectively 
by a compact discrete subgroup of it. We argued that for the two solvable cases we found, there are no 
associated compact spaces. However, all the cases of nilpotent Lie algebras give rise to certain compact 
nilmanifolds. These nilmanifolds can be formally quantized via Weyl quantization. 
%In particular, we determined 
%the corresponding star product in each case. 
%Therefore these quantized nilmanifolds correspond to compact NC 
%spaces ${\cal M}_{\ast}$, which are solutions of the IKKT model.
Although it cannot be argued at this stage that these compact manifolds are solutions of the IKKT model as 
well, it would be interesting to investigate the compactification of the model on them along the lines 
of \cite{Connes:1997cr}.

%Then the full 10-dimensional solution takes the form $\R^4_{\theta}\times {\cal M}_{\ast}$ 
%and it corresponds to a compactification with a quantized nilmanifold as internal space. This point of view 
%is very plausible, since similar compactifications have been identified directly in the context of type II 
%string theories \cite{LopesCardoso:2002hd,Dall'Agata:2005ff,Hull:2005hk,Grana:2006kf}, where they are 
%usually dubbed ``twisted tori''. In string vacua these twisted tori carry fluxes associated to structure 
%constants of a group. Here these fluxes are directly encoded in the geometry defining the corresponding NC 
%space. The above picture is similar to the one of \cite{Connes:1997cr} and provides support to the conjecture that the IKKT model is related to the type IIB 
%superstring theory. 

\bigskip

\paragraph{Acknowledgements.} Useful discussions with L. Jonke, A. Kehagias and H. Steinacker are gratefully 
acknowledged. This work was partially supported by the SFB-Transregio TR33
"The Dark Universe" (Deutsche Forschungsgemeinschaft) and
the European Union 7th network program "Unification in the
LHC era" (PITN-GA-2009-237920).

\bigskip

\vspace{20pt}

\appendix

{\Large{\textbf{Appendix}}}

\section{Definitions on Lie algebras}

Lie algebras are classified according to their properties in simple, semisimple, abelian, nilpotent 
and solvable. In this brief appendix let us collect some useful definitions, which appear often in the main 
text. A standard reference is \cite{Fuchs:1997jv}.
%Evidently, a Lie algebra is abelian if its Lie bracket vanishes for any two of its elements. 
%A Lie algebra is solvable 

Let us denote a Lie algebra by $\cal A$ and its generators by $X_a$. These generators satisfy the commutation 
relations
\be [X_a,X_b]=f_{ab}^cX_c, \ee 
where $f_{ab}^c$ are the structure constants. Knowledge of the structure constants is enough to determine the 
Killing form $g_{ab}$ according to the formula
\be  g_{ab}=f_{ac}^df_{bd}^c. \ee
According to Cartan's criterion a Lie algebra is semisimple if and only if its Killing form is non-degenerate. 
Accordingly, the Killing form of a nilpotent Lie algebra vanishes identically.

Given a Lie algebra one may search for its invariants, i.e. functions of its generators which commute with all 
the generators. The most important of these invariants is the quadratic Casimir operator, which we denote as 
\be C^{(2)}({\cal A})=\Omega^{ab}X_aX_b, \ee
with $\Omega^{ab}$ the elements of a symmetric matrix. If the matrix $\Omega^{ab}$ is invertible, then one may 
form its inverse $\Omega_{ab}$, which is symmetric, non-degenerate and invariant under the adjoint action of 
the corresponding group. In other words, $\Omega_{ab}$ is a metric on the corresponding group manifold. In 
fact, for semisimple algebras $\Omega_{ab}$ is proportional to the Killing form. This is no longer true 
for non-semisimple ones.

Let us define the derived algebra ${\cal A}'$ of a Lie algebra ${\cal A}$ as 
\be {\cal A}'=[\cal A,\cal A]. \ee 
Moreover, let us introduce two generalizations of the derived algebra and in particular the so-called upper 
central series or derived series of $\cal A$, defined as
\be {\cal A}^{\{i\}} = [{\cal A}^{\{i-1\}},{\cal A}^{\{i-1\}}], \quad i\ge 2 \ee
and the lower central series, defined as 
\be {\cal A}_{\{i\}} = [{\cal A},{\cal A}_{\{i-1\}} ], \quad i\ge 2.
\ee In the above iterative definitions it holds that ${\cal A}_{\{1\}}={\cal A}^{\{1\}}={\cal A}'$. Then we have the following 
two definitions:
\begin{itemize}
 \item A Lie algebra is called solvable if its derived series becomes zero after a finite number of 
steps, i.e. $\exists\ i_0$, such that ${\cal A}^{\{i_0\}}={0}$.
\item A Lie algebra is called nilpotent of step $i_0$ if 
its lower central series becomes zero after a finite number of steps, i.e. $\exists\ i_0$, such that 
${\cal A}_{\{i_0\}}={0}$.
\end{itemize}
 Clearly nilpotency implies solvability.

\end{document}